\documentclass[aoas,nameyear,MSNbibl,dvips]{arximspdf}

\doi{10.1214/10-AOAS398I}
\volume{5}
\issue{1}
\pubyear{2011}
\firstpage{45}
\lastpage{46}
\referstodoi{10.1214/10-AOAS398}

\begin{document}
\begin{frontmatter}

\title{Discussion of:  A statistical analysis of multiple temperature proxies: Are
reconstructions of surface temperatures over the last\\ 1000 years reliable?\thanksref{AUT1}}

\runtitle{Discussion}

\pdftitle{Discussion on A statistical analysis of multiple temperature proxies: Are
reconstructions of surface temperatures over the last 1000 years reliable?
by B. B. McShane and A. J. Wyner}

\begin{aug}

\author{\fnms{L. Mark} \snm{Berliner}\corref{}\ead[label=e1]{mb@stat.osu.edu}}

\thankstext{AUT1}{Supported by NSF Grant ATM-0724403.}

\runauthor{L. M. Berliner}

\affiliation{The Ohio State University}

\address{Department of Statistics\\
The Ohio State University\\
1958 Neil Avenue\\
Columbus, Ohio 43210-1247\\
USA\\
\printead{e1}} 
\end{aug}

\received{\smonth{9} \syear{2010}}
\revised{\smonth{9} \syear{2010}}



\end{frontmatter}

I join the authors in expressing dissatisfaction with some
paleoclimate analyses.  I endorse
their claim that there has been underestimation of uncertainty in
paleoclimate studies. The implication  that
additional participation of the statistics community is needed is
undeniable. However, our priorities should be to contribute rich
statistical analyses that (i) model the processes and data
and (ii) offer useful information regarding the issues of climate change.
If achieving these goals requires that we do not continue with questionable
assumptions, nor merely offer small fixes to previous approaches,
nor participate in uncritical debates, so be it.

The authors note that it is common to
assume that proxy observations are linearly related to
climate variables and they proceed with this assumption.
This seems untenable to me (for an extreme example see the
Yellow River data in Figure 6). Even if linearity is plausible,
lumping all spatial-temporally distributed data of various types,
qualities, and degrees of relationship to climate variables into
a variance--covariance based summarization (principal components or
EOFs) with no underlying analysis gives me pause. I am not surprised
by difficulties in then extracting usable information. Performing
various tests and analyses
based on these reductions seems of little interest; indeed, it seems
to me that they serve as a distraction.

Leaping ahead, though I strongly endorse the
application of Bayesian analysis in this context, the concerns
of the previous paragraph remain active regarding the Bayesian
analysis in this article. Indeed, much like other
analyses, the assumption is that regressing onto principal components
with coefficients constant in time captures enough of the structure of
the process to base the modeling on a stationary, AR(2) model. This
places a reliance on the principal components that I find highly
questionable. At a minimum, it seems to me that using spatially
distributed and proxy dependent regression coefficients should be
considered. Such an approach is closer to what I would call a ``modern
Bayesian analysis.''

To provide perspective I return to my remark regarding ``uncritical
debates.''  The overarching conclusion of the authors seems to be that
warming is real, but that the specifics of the rapid uptake associated
with the ``hockey stick'' is not supported by the data. First, the
claim is not unequivocal.  As mentioned, I find that there are serious
concerns with the analyses. In addition, we know that there are many
controllers of climate; indeed, we know humans have
contributed to some of these controllers. Hence, these analyses
have ignored data.  There is no use of atmospheric CO$_2$ data or
solar data nor adjustment for climate variations associated with the
El Ni\~{n}o-Southern Oscillation, the Pacific Decadal Oscillation,
etc. What should we make of results of any analyses that seek to use
high-temperature, high-CO$_2$-level temperatures to back-cast
temperatures with no adjustment for CO$_2$?  To me, not much, given that I
do not believe the principal components can account for all the known
and unknown sources of variation and nonstationarity. [For a very
simple example of how we might account for such things,
see Berliner and Kim (\citeyear{BK2008}).]

Second, even if we
accept the ``no-hockey'' conclusion, is it critical to the climate
policy debate? I believe not, though I acknowledge that some policy
makers and a portion of the general public do not understand the
issues.
The problem of anthropogenic
climate change cannot be settled by a purely statistical argument. We
can have no controlled experiment with a series of exchangeable
Earths randomly assigned to various forcing levels to enable
traditional statistical studies of causation. (The use of large-scale
climate system models can be viewed as a surrogate, though we need to
better assess this.) Rather, the issue
involves the combination of statistical analyses {\it and}, rather than
versus, climate science.
Combination of information, such as that in Figure 15 along
with climate model data based on anthropogenic and natural forcings
versus only natural forcings along with uncertainty quantification
constitute the basis for contributing to the climate change problem.

\def\bibname{Reference}

\printaddresses

\end{document}